\documentclass[prl,twocolumn,amsmath,amssymb,superscriptaddress,showpacs]{revtex4}

\usepackage{graphicx}
\usepackage{epsfig}
\usepackage{epstopdf}
\begin{document}


\title{Electron and optical phonon temperatures in electrically biased graphene}

\author{St\'ephane Berciaud}
\email{sb2840@columbia.edu}
\affiliation{Departments of Physics and Electrical Engineering, Columbia University, New York, NY 10027}
\affiliation{Department of Chemistry, Columbia University, New York, NY 10027}
\author{Melinda Y. Han}
\affiliation{Department of Applied Physics and Applied
Mathematics, Columbia University New York, NY 10027}
\author{Louis E. Brus}
\affiliation{Department of Chemistry, Columbia University, New York, NY 10027}
\author{Philip Kim}
\affiliation{Department of Physics, Columbia University, New York, NY 10027}
\author{Tony F. Heinz}
\email{tony.heinz@columbia.edu}
\affiliation{Departments of Physics and Electrical Engineering, Columbia University, New York, NY 10027}


\begin{abstract}
We examine the intrinsic energy dissipation steps in electrically biased graphene channels. By combining in-situ measurements of the spontaneous optical emission with a Raman spectroscopy study of the graphene sample under conditions of current flow, we obtain independent information on the energy distribution of the electrons and phonons. The electrons and holes contributing to light emission are found to obey a thermal distribution, with temperatures in excess of  $1500$~K in the regime of current saturation. The zone-center optical phonons are also highly excited and are found to be in equilibrium with the electrons. For a given optical phonon temperature, the anharmonic downshift of the Raman G-mode is smaller than expected under equilibrium conditions, suggesting that the electrons and high-energy optical phonons are not fully equilibrated with all of the phonon modes.
\end{abstract}

\maketitle

The charge transport properties of the two-dimensional system of single-layer graphene have been the subject of much recent interest~\cite{Geim07}. Graphene is characterized by remarkably large room-temperature carrier mobilities ($>100 ~000$~cm$^2$V$^{-1}$s$^{-1}$)~\cite{Bolotin08}, as well as by its ability to sustain high current densities ($>10^8$~A/cm$^{2}$)~\cite{Meric08,Barreiro09,Freitag09,Chae10}. While the physical processes defining the carrier mobility and scattering at low fields have been probed experimentally~\cite{Chen08}, relatively little is known about the underlying physics of high-field transport and, in particular, the mechanisms responsible for the observed current saturation with increasing bias~\cite{Meric08,Barreiro09,Freitag09,Chae10}. Recent investigations indicate a critical role for coupling of energetic carriers with the high-energy optical phonons~\cite{Barreiro09}. The distribution of energy within and among the charge carriers, the strongly coupled optical phonons, and the other phonons in graphene has, however, not yet been established. Knowledge of these quantities is fundamental to establishing the origin of current saturation. This is a subject not only of great inherent interest, but also of importance for emerging applications of graphene in high-performance field-effect transistors~\cite{Xia10,Lin10}.  Non-equilibrium behavior in the saturation regime, including an over-population of the strongly coupled optical phonon modes, has previously been invoked to explain the nature of high-field electrical transport in carbon nanotubes~\cite{Pop05,Lazzeri06,Krupke08,Steiner09}.

In this paper, we present the results of \textit{in-situ} measurements of the energy distribution in key degrees of freedom of graphene under high electrical bias. We introduce a new method for analysis of the energy distribution of the charge carriers. The approach is based on characterization of light emission from the biased graphene.  We find that over an energy range of $0.7-1.3$~eV, the carriers can be described by a thermal distribution, with temperatures $T_{el}^{~}$ reaching values as high as 2000~K before device failure. In addition, we apply Raman spectroscopy to establish the population of the zone-center (G-mode) optical phonons $T_{op}^{~}$. This temperature matches, to experimental accuracy, that of the electrons. 
Based on the frequency shift of the G-mode phonons in the Raman measurements, we can also learn about the population of the lower-energy anharmonically coupled modes.  They too are significantly heated, but the observed frequency shift is smaller than the value expected under equilibrium conditions, indicating that the strongly coupled optical phonons may not be in full equilibrium with the other phonon modes of the graphene lattice.  This latter observation is compatible with the trend reported in the recent work of Chae~\textit{et al.}~\cite{Chae10, ChaeComment}.  

In our experiments, graphene monolayers were prepared by mechanical exfoliation ~\cite{Geim07} of kish graphite.  The layers were deposited onto highly doped Si substrates covered with a $285\pm5$~nm layer of thermal SiO$_2$, used as a gate dielectric. Cr/Au electrical contacts were fabricated using standard electron-beam lithography followed by metal evaporation. Graphene channels having nearly rectangular shape were chosen for our devices to avoid any additional etching processes.  The monolayer thickness of the graphene samples was confirmed by analysis of the Raman 2D mode~\cite{Ferrari06}. The graphene devices were held at room temperature and under vacuum (at pressures below 10$^{-7}$ Torr). The optical radiation emitted by electrically biased graphene was collected in a backscattering geometry using an achromatic objective lens ($\times$50, NA=0.5).  The light was then dispersed onto a charged-coupled device array by a single-pass optical spectrometer. The same experimental configuration was used to record Raman scattering spectra. In that case, a~2.33~eV laser beam was focused onto a 2~$\mu$m diameter spot at the center of the graphene channel. The laser power absorbed by graphene~\cite{Mak08,Yoon09} was very low ($<2$~kW/cm$^2$ ) to avoid any laser-induced heating of the sample~\cite{Calizo07}. In order to normalize the experimental data, we recorded the spectral response of our system using a calibrated tungsten-halogen lamp emitting the spectrum of a black body at a temperature of 3400~K.

\begin{figure}[!t]
\includegraphics[scale=0.70]{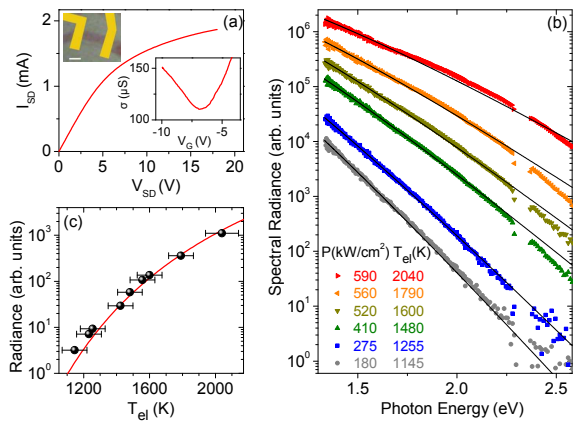}
\caption{(color online) (a) Source-drain current-voltage characteristics at zero gate bias of the 3.6$\times$1.6$~\mu$m$^2$ graphene channel shown in the top inset (the scale bar is 2~$\mu$m). The lower inset shows the back-gate dependence of the channel conductivity ($\sigma$). (b) Spectral radiance of the same graphene sample as a function of the electrical power per unit area $P$ dissipated in the channel (symbols). The spectra are fit to Planck's law (solid lines). In the fits, the thickness of the SiO$_2$ layer was fixed at $280\pm2$~nm. The dips around 2.33~eV are artifacts arising from the notch filter used for the Raman measurements [Fig.~\ref{figRaman1}].~(c) Radiance for photon energies from 1.33~eV$-$2.59~eV as a function of the electronic temperature ($T_{el}^{~}$) deduced from the fits in (b). The solid line is a fit based on Planck's law.}
\label{figBBR} 
\end{figure}

Figure~\ref{figBBR}a displays the source-drain current-voltage ($I_{SD}^{~}-V_{SD}^{~}$) characteristics of a typical graphene channel measured at zero gate bias ($V_G^{}$). The minimum conductivity point was found at $V_G^{}=-7$~V, which suggests a residual charge density in the range $10^{11} \rm\:cm^{-2}-10^{12} \rm\: cm^{-2}$ at $V_G^{}=0$~V . In agreement with previous transport measurements~\cite{Meric08,Barreiro09,Freitag09,Chae10}, we find that the current through the channel, I$_{SD}^{~}$, shows saturation behavior in the high-bias regime [Fig.~\ref{figBBR}a]. During the measurements, we observed slight drifts in I$_{SD}^{~}$, which are attributed to bias- and laser-induced shifts of the Fermi level. Thus, in our study of the electron and phonon populations, we present our results as a function of $P$, the average dissipated electrical power per unit area of the channel. 

For sufficiently high $P$, optical emission from graphene can be readily observed \textit{even in the absence of optical excitation}. The raw optical emission spectra were normalized by the spectral response of our detection system and, importantly, by the computed energy-dependent enhancement factor arising from multiple reflections in the graphene/SiO$_2$/Si  multilayer system~\cite{Yoon09}. The corrected spectra are shown in figure~\ref{figBBR}(b) as a function of $P$. We observe a rise of the spectrally integrated intensity by nearly three orders of magnitude [Fig.~\ref{figBBR}(c)], while $P$ only increases by a factor of $\approx3$. This enhancement is associated with pronounced blue-shifts of the emission spectra with increasing $P$. We attribute the optical emission to bias-induced spontaneous emission from energetic charge carriers in the graphene channel. In order to investigate this phenomenon quantitatively, we consider the case of thermalized charge carriers and fit the corrected spectra using Planck's law for the spectral radiance $u$ of a \textit{grey} body, namely:  
	\[u(E,T_{el}^{~})=\epsilon(E)~\frac{2E^3}{h^2c^2}\left[\exp\left(\frac{E}{k_B^{~} T_{el}^{~}}\right)-1\right]^{-1}
.\] In this expression, $h$ is Planck's constant, $c$ is the speed of light in vacuum, $E$ is the energy of the emitted photon, and $T_{el}^{~}$ is the temperature of the  electronic system.  Following recent absorption measurements~\cite{Mak08}, we assume that the emissivity $\epsilon(E)$ of a graphene monolayer is constant over the energy range studied here ($\epsilon\approx2.3\%$~\cite{Mak08}). We note that since optical transitions in the visible spectral range correspond to electronic excitations in the system, we associate the inferred temperature with that of the electronic system of graphene.

The agreement between our data and this model, which uses no adjustable parameter, is remarkably good [Fig.~\ref{figBBR}(b)]. The extracted $T_{el}$ range from 1145~K at the lower end of our detection sensitivity up to 2040~K for currents near the point of device failure. Also, the radiance integrated for photon energies from 1.33~eV$-$2.59~eV scales well with the corresponding integration of Planck's law [Fig.~\ref{figBBR}(c)].  Optical measurements over this energy range reflect the behavior of electron and hole populations in graphene each with half the photon energy, i.e., from $\sim 0.7 - 1.3$~eV. Our results demonstrate that under high electrical bias these electrons and holes thermalize rapidly and can attain very high temperatures. As expected for the relevant energy range, the electron and hole distributions do not exhibit any observable sign of their drift velocity from the bias field. 

\begin{figure}[!t]
\includegraphics[scale=0.8]{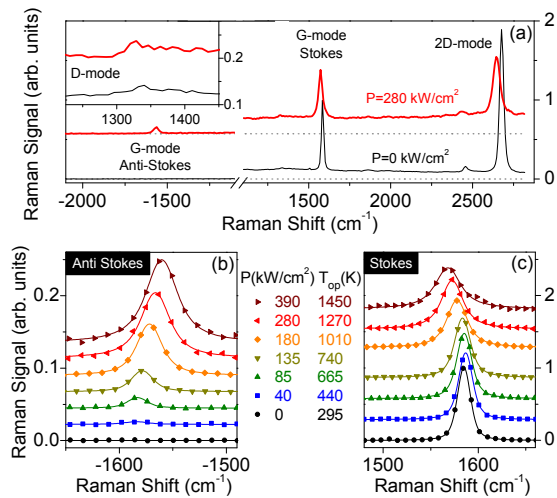}
\caption{(color online) (a) Broad band Raman spectra of the same graphene channel as in Fig.~\ref{figBBR} under zero- and high- source-drain biases. The inset is a close-up showing the weak D-mode. (b) Anti-Stokes and (c) Stokes Raman G-modes (symbols). The solid lines are Voigt fits. The spectra are vertically offset for clarity. All spectra were acquired with the same integration time. The dissipated electrical power per unit area ($P$) and the optical phonon temperature ($T_{op}^{~}$) are indicated.}
\label{figRaman1} 
\end{figure}

We now examine the coupling between the electronic system and high-energy optical phonons.
As shown in Fig.~\ref{figRaman1}, the Raman spectrum of electrically biased graphene is markedly different from a reference spectrum taken at $V_{sd}^{}=0$~V. When $P$ increases, the thermal emission from the electrons produces an increasing background in the Raman spectra, which eventually overcomes the Raman signal for $P>500$~kw/cm$^2$. We observe that the integrated intensity of the anti-Stokes G-mode increases strongly relative to that of the Stokes counterpart. This effect is accompanied with a downshift and a broadening of both G- and 2D- Raman features. Finally, the integrated intensity of the defect-related D-mode is less than $5\%$ of that of the G-mode [Fig.~\ref{figRaman1}(a), inset], indicating that the high structural quality of our samples is preserved. We did not observe any significant enhancement of the D-mode up to near the device breakdown threshold, where an enhanced and broadened D-mode ultimately develops. This trend was observed on several graphene channels, with similar breakdown current densities of $\sim4\times$10$^8$~A/cm$^2$.

The absolute phonon population for the zone-center optical phonons can be derived from our measurements of the Raman G-mode. Indeed, the Stokes signal is proportional to $1+n_G^{}(T_{op}^{~})$, where $n_G^{}$ is the G-mode population with an effective temperature $T_{op}^{~}$ following from the Bose-Einstein distribution, while the anti-Stokes signal scales as $n_G^{}(T_{op}^{~})$. Thus the \textit{measured} integrated intensity ratio of the anti-Stokes and Stokes features [Fig.~\ref{figRaman1}(b,c)] can be written as 
	\[\frac{I_{aS}^{~}}{I_{S}^{~}}=C\times \exp\left(-\frac{\hbar \omega_{G}^{~}}{k_{B}^{~} T_{op}^{~}}\right)
,\]
where $\hbar \omega_G^{~}$ is the G-phonon energy ($\approx$195meV), $k_{B}^{~}$ is the Boltzmann constant and $C$, is a numerical factor.

Our physical conclusions depend critically on an accurate measurement of $C$~\cite{ChaeComment}. 
Thus, we measured $I_{aS}^{~}/I_{S}^{~}$ at room temperature on an unbiased sample, using a very low laser power. We obtained a value of $C=0.85\pm0.15$, which we have used to extract $T_{op}$. The same experimental factor of $C=0.85$ is also obtained if we assume a $E^4$ dependence of the Raman signal on the scattered photon energy~\cite{Kuzmany98} and properly take into account the distinct spectral responses of our setup and interference enhancements at the energies of the Stokes and anti-Stokes scattered photons~\cite{Yoon09}. 

\begin{figure}[!t]
\includegraphics[scale=0.5]{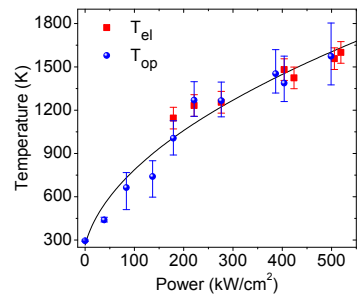}
\caption{(color online) Electronic ($T_{el}^{~}$, squares) and optical phonon ($T_{op}^{~}$, circles) temperatures as a function of the dissipated electrical power. The dashed line is a guide to the eye, based on a scaling of the temperature as $\sqrt{P}$.}
\label{figTGTe} 
\end{figure}

The values of $T_{op}$ can also be derived directly from the intensity $I_S^{~}$ of the Stokes Raman signal. Since the optical phonon population $n\ll1$ in the absence of current flow, we obtain $I_S^{~}(T_{op}^{})/I_S^{0}=1+n_G^{}(T_{op}^{})$, where $I_S^{0}$ is the Stokes intensity for $V_{sd}^{}=0$. The extracted values of $T_{op}^{~}$ using this approach are in good agreement with those determined from the ratio $I_{aS}^{~}/I_{S}^{~}$. In the following, $T_{op}^{~}$ will be taken as the mean of these two independent determinations. We find values of $T_{op}^{~}$ from 440~K up to 1570~K as $P$ increases from 40~kW/cm$^2$ to 500~kW/cm$^2$ [Fig.~\ref{figRaman1} and Fig.~\ref{figTGTe}].

Our results allow us to perform a direct comparison of $T_{el}^{~}$ and $T_{op}^{~}$ at a function of $P$. Within experimental accuracy, we find that these temperatures are identical [Fig.~\ref{figTGTe}].  At high $P$, the increase in temperature with increasing power saturates. This behavior cannot be explained by losses through the observed radiation. Even at the highest temperatures measured, the radiated power given by the Stefan-Boltzman law remains a negligible fraction ($\sim4.10^{-6}$) of $P$.  The observed saturation presumably reflects the rate of perpendicular heat flow to the underlying substrate~\cite{Freitag09}.

\begin{figure}[!t]
\includegraphics[scale=0.6]{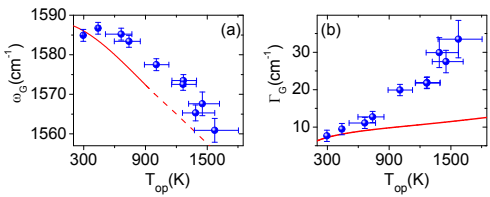}
\caption{(color online) (a) Frequency ($\omega_{G}^{~}$) and (b) Lorentzian full width at half maximum ($\Gamma_{G}^{~}$) of the G-mode feature  as a function of $T_{op}^{~}$. Data are extracted from Voigt fits, taking into account a constant Gaussian contribution (13~cm$^{-1}$) due to our setup spectral resolution. 
The thick solid lines in (a,b) are the theoretically predicted temperature dependences of $\omega_{G}^{~}$ and $\Gamma_{G}^{~}$ under equilibrium conditions~\cite{Bonini07}. The dashed line in (a) is a linear extrapolation of the results in ref.~\cite{Bonini07}, neglecting higher-order anharmonic terms. To compute the solid line in (b), we considered a residual charge density of $10^{12}$cm$^{-2}$ and an anharmonic decay rate of 2.5~cm$^{-1}$ at $T=300\rm~K$~\cite{Yan09}.}
\label{figRaman2} 
\end{figure}

Is this subset of high-energy optical phonons in equilibrium with the lower-energy phonon modes in graphene? This question can be addressed by studying the spectral line shape of the Raman G-mode under electrical bias [Fig. \ref{figRaman2}(a,b)]. The Raman features are known to undergo a frequency downshift with increasing temperature as a result of anharmonic coupling~\cite{Bonini07}. Under equilibrium heating, the G-mode downshift reflects primarily the \textit{anharmonic} temperature of these secondary phonons into which G-mode phonons decay.

For moderate temperatures, the downshift of the G-mode may be approximated as a linear function of the equilibrium temperature of the system ~\cite{Calizo07}.  Such an approximation becomes questionable in the saturation regime where high temperatures may be reached~\cite{Freitag09}.  For our analysis, we have therefore made use of the theoretical predictions from Bonini~\textit{et al.}~\cite{Bonini07}. Although the theory has not been tested experimentally at high temperatures, the results are compatible with available low-temperature data ($100-400 \rm\: K$)~\cite{Calizo07}. We have used the theoretical results to estimate the downshift of $\omega_{G,eq}^{~}$ expected if the system were in full thermal equilibrium  at a temperature $T_{op}^{~}$. As shown in Fig.~\ref{figRaman2}~(a), the predicted $\omega_{G,eq}^{~}$ are somewhat lower than the measured values of $\omega_G^{~}$. This indicates that the strongly coupled hot electrons and phonons are not in full equilibrium with the lower energy phonons. For the highest temperature regime($\text{T}>900~\text{K}$), we have simply extrapolated the results of Bonini~\textit{et al.}.  This procedure  almost certainly underestimates the rate of downshift, since it neglects contributions from higher-order anharmonic terms, which would imply a larger departure from equilibrium.  Nevertheless, we should stress that the observation of frequency downshifts as large as $25 \rm~cm^{-1}$ for $\omega_G^{~}$ implies that the secondary phonon modes are significantly heated and may reach temperatures on the order of $1000\rm\:K$.  We may infer that heating of all accessible degrees of freedom in the graphene sample is significant, although complete thermal equilibrium does not appear to be attained.

A further interesting point concerns the bias-induced broadening of the Raman features. This effect is completely reversible and can lead to an increase by as much as a factor of $\sim5$ for the width of the G-mode ($\Gamma_{G}^{~}$) [Fig.~\ref{figRaman2}~(b)]. This width arises from two intrinsic components: coupling between G-mode phonons and resonant electron-hole pairs (i.e., Landau damping), and from the anharmonic coupling described above~\cite{Bonini07}. The first contribution would actually imply a decrease in $\Gamma_{G}^{~}$ for high temperatures.  The anharmonic coupling would lead to broadening of the G-mode with increasing temperature.  However, as shown in Fig.~\ref{figRaman2}~(b), the magnitude of the broadening far exceeds that expected from the lifetime of the lowest-order anharmonic processes~\cite{Bonini07}.  We believe that the observed broadening may result from enhancement of the quasi-elastic scattering rate, as well as possible contributions from higher-order anharmonic terms that become significant at high temperatures.

Having examined the behavior of zone-center (G-mode) optical phonons, we finally address that of the zone-edge optical phonons that give rise to the (symmetry-forbidden) D-mode and the two-phonon (symmetry-allowed) 2D-mode~\cite{Maultzsch04,Basko09}. Analysis of the D-mode is difficult since the signal is weak and strongly influenced by the defect density, which may in itself be altered by temperature.  Extracting phonon populations from the 2D-mode is also challenging. The 2D-mode involves electronically resonant intervalley scattering with large-momentum phonons~\cite{Maultzsch04,Basko09}. Thus, as opposed to that of the G-mode, its intensity will reflect the competition between electron-phonon and electron-electron scattering processes.  According to the model of~\cite{Basko09}, we have $I_{2D}^{~}\propto \left( {1+\gamma_{el}/\gamma_{ph}}\right)^{-2}(1+n_D^{})^2$, where $n_D^{}$ is the zone-edge optical phonon population at a temperature $T_{op}^{}=T_{el}^{}$, and $\gamma_{ph}$  and $\gamma_{el}$ are, respectively, the electron-phonon  and electron-electron scattering rates. A strong increase in $\gamma_{el}^{}$ is presumably the origin of the observed decrease of $I_{2D}^{}$ with increasing optical phonon temperature [Fig.~\ref{figRaman1} (a)].

In conclusion we have performed an \textit{in situ} study of the electron \textit{and} and phonon populations in  graphene field-effect transistors operating in the high-bias regime. Electrons and holes with energies from $0.7-1.3$~eV are found to be in a thermal equilibrium distribution and to achieve temperatures as high as 2000 K without device failure. Direct measurement of the zone-center optical phonon population shows that they are in thermal equilibrium with the electronic excitations of the system.  The lower-energy phonons are also substantially heated, but appear to be characterized by a somewhat lower temperature than that of the electrons and optical phonons. This strong coupling between electrons and optical phonons is a hallmark of graphitic materials and is assuredly the main intrinsic mechanism responsible for current saturation in graphene field-effect transistors.  As regards the possible role of SiO$_2$ surface phonons in the microscopic processes underlying current saturation~\cite{Meric08,Barreiro09,Chen08,Perebeinos10}, our results demonstrate that, although extrinsic contributions from such modes cannot be ruled out, large populations of optical phonons (up to $n\sim0.5$) are present in the high-bias regime.  

We acknowledge support from Office of Naval Research MURI program and the DARPA CERA program.

\end{document}